\documentclass[fleqn,usenatbib]{mnras}

\usepackage[T1]{fontenc}
\usepackage{ae,aecompl}

\DeclareRobustCommand{\VAN}[3]{#2}
\let\VANthebibliography\thebibliography
\def\thebibliography{\DeclareRobustCommand{\VAN}[3]{##3}\VANthebibliography}

\usepackage{mathtools}
\usepackage[varg]{txfonts}
\usepackage{color}
\usepackage{natbib}
\usepackage{multirow}   
\usepackage{gensymb}    
\usepackage{graphicx}
\usepackage{url}
\usepackage{indentfirst}
\usepackage[normalem]{ulem}     
\usepackage{placeins}
\usepackage{tablefootnote}
\usepackage{amsmath}	
\usepackage{amssymb}	
\hypersetup{colorlinks=true,allcolors=blue}

\title[Low-frequency break detections in PDS of GX 339-4]{Detection of low-frequency breaks in power density spectrum of GX 339-4 in faint low/hard state observations using AstroSat data }

\author[Husain et al. 2021]{
Nazma Husain,$^{1}$\thanks{E-mail: nazmahusain2017@gmail.com}
Ranjeev Misra,$^{2}$
Somasri Sen$^{1}$
\\
$^{1}$Department of Physics, Jamia Millia Islamia, New Delhi-110025, India\\
$^{2}$Inter-University Centre for Astronomy and Astrophysics, Pune-411007, India\\
}

\date{Accepted XXX. Received YYY; in original form ZZZ}

\pubyear{2021}
\begin{document}
\label{firstpage}
\maketitle

\begin{abstract}
 
\noindent We present the spectro-timing analysis of the black hole binary system GX 339-4 using AstroSat data extracted at the beginning of outbursts in 2017 and 2019. The joint spectral fitting of LAXPC and SXT data revealed that the source was in a faint low/hard state for both 2017 and 2019 observations with a nearly equal photon index of $\sim$1.57 and $\sim$1.58 and with Eddington ratio, $L/L_{Edd}$, of 0.0011 and 0.0046 respectively. The addition of a reflection component into the spectral modelling improved the fit ($\Delta\chi^2$ $\approx~6$ for 2017 and $\Delta\chi^2$ $\approx~7$ for 2019), pointing towards the presence of weak reflection features arising due to irradiation of the accretion disk. The power density spectrum (PDS) consisted of strong band-limited noise with a break at low frequencies, described with a combination of few zero-centered Lorentzian. The fitting revealed a low-frequency break at $\sim$6~mHz for 2017 and at $\sim$11~mHz for 2019 observation, whose detection is validated by results from independent detectors (LAXPCs and SXT). The break frequency is roughly consistent with results obtained from earlier observations that showed an evolution of the frequency with flux, which is in accordance with the truncated disk model. Associating the break frequency with viscous time scale of the accretion disk, we estimated a truncation radius of $\sim$93 and $\sim$61 gravitational radius for 2017 and 2019 observation, respectively.

\vspace{0.3cm}
\end{abstract}


\section{Introduction}

A Black hole X-ray binary system (BHXB) is a system consisting of a black hole and a companion star orbiting about their common centre of mass. The black hole due to its  gravitational field accretes matter from the orbiting star via Roche-lobe overflow mechanism or stellar wind depending upon the mass of  the  companion  star.  Infalling  matter  has  significant angular  momentum  which  allows for the  formation  of  an accretion  disk around  the  black  hole. This disk is luminous in X-rays in  the  vicinity  of  the  compact  object. Another component of flow exists near the compact object which is hotter and optically thinner than the disk and is responsible for the hard X-ray emission known as hot inner flow \citep{sunyaev1979hard,thorne1975cygnus}. The exact geometry of this flow is still under discussion. Till now, many Galactic black hole binaries have been detected and analysed amongst which GX 339-4 (V821 Ara) remains one of the most extensively  observed  low  mass  BHXB  system  due  to  its  variety  of spectral states and complex outburst profile. The accretion of matter in this source is dominated by Roche-lobe overflow mechanism due to its low mass companion star \citep{heida2017mass}. Since its discovery in 1973 by satellite OSO-7 \citep{markert1973observations},  frequent outbursts have been observed every 2-3 years consistent with its transient nature. During these outbursts, GX 339-4 manifests spectral and  temporal  transitions  which  may  depend  on  the  geometry  and luminosity of the two accretion flows \citep{gilfanov2010spectral}.\\

Different  locations  on  the Hardness  Intensity  Diagrams  (HID)  describe different states of the binary system, with each state exhibiting different spectral and timing properties. The distinction between the states is explained in \cite{belloni2006states, belloni2010states}. Many black hole transients including GX 339-4 follow a q-shaped HID \citep{homan2005,dunn2010} during their complete outburst. In beginning of the outburst, the source starts in the low/hard state (LHS) with a dominating power-law component (photon index, $\Gamma$ $<2$) originating from the inner hot flow. As the luminosity increases, the source transits to the Hard Intermediate State (HIMS) which is a softer state ($\Gamma$ >2) and has an increased amount of disk contribution to the spectrum but with little change in its timing properties as compared to the low/hard state. Further along the outburst, the source evolves into the Soft Intermediate State (SIMS) with further softening of the spectrum and significant change in timing properties. From the SIMS the source then transits to the Soft state (SS) which has dominating disk emission modelled by a disk multi-color black body spectrum. Similar trend is followed during the decline of the outburst but in reverse order, hence putting the source back into the low/hard state, completing the q-shaped HID. \\

One of the ways to understand the accretion flow in strong gravitational field is to study the timing variability of the flow. A common technique for X-ray timing analysis is to study the Fourier transform of the X-ray flux, also expressed in terms of Power Density Spectrum (PDS) \citep{van1989fourier}. With changing states the PDS has been observed to show different components. In low/hard state the PDS consists of a strong band-limited noise (BLN) continuum \citep{oda1971BLN} with a break at low-frequency \citep{titarchuk2007power} and sometimes accompanied with a low-frequency Quasi-Periodic Oscillation (QPO). The PDS in this state is usually fitted with a few broad Lorentzians or a broken power-law model \citep{belloni2002unified,stiele2017nustar,nowak2000there}. As the source rises in luminosity and moves towards softer states the rms variability reduces and  prominent type-B ($\sim$5-6~Hz) and type-C ($\sim$0.1-15~Hz) QPOs are observed, see for \textit{e.g.} \cite{casella2005abc}. \\

The band-limited noise detected in low/hard state has a characteristic low-frequency break ($\nu_{brk}$) which has been observed to vary during the cycle of outburst. In $\nu$~$P_\nu$ representation, where $P_\nu$ is the normalized power, the break characterises the PDS such that below this the power spectrum is a power-law of index -1 and above the break it takes the shape of a flattened power-law \citep{miyamoto1992,belloni1990a}. This low-frequency break has been previously detected in the PDS of GX 339-4 in several studies. For \textit{e.g.}, \cite{nowak1999low} studied the low luminosity RXTE observations of GX 339-4, and detected a break frequency at $\sim$30~mHz. \cite{ford1998measurement} detected a break at 70~mHz for one of the hard state observations of 1997 outburst for GX 339-4. \cite{migliari2005} also studied this outburst and detected the low-frequency break in frequency range 37 to 107~mHz for its hard state observations.  \cite{nandi2012accretion} performed a detailed analysis of 2010-2011 outburst of this source and observed the break frequency in different states amongst which the only hard state observation exhibited a break at 880~mHz. Similarly, \cite{stiele2015energy} analysed two hard state observations of GX 339-4 extracted in 2004 and 2009 and found low-frequency breaks at 44.7 and 8.7 mHz respectively. Also,  break frequencies in the low luminosity state were detected using NuSTAR data for 2015 outburst by \cite{stiele2017nustar} in range 30-70mHz.

In addition to these detections, \cite{wijnands1999break} studied the aperiodic variability of multiple BHXBs including GX 339-4, Cyg X-1, XTE J1748-288, GRO J1655-40 and few others in hard state. They detected the break for GX 339-4 in the range 40-64~mHz. They have also observed a positive correlation between the  low-frequency QPO ($\nu_{QPO}$) and the low-frequency break ($\nu_{brk}$) such that $\nu_{QPO}$~$\sim5$~$\nu_{brk}$, which implies that both band-limited noise and low-frequency QPO might have a common physical origin.  Furthermore, \cite{plant2015truncated} showed that as GX 339-4 progresses through the hard state, the low-frequency break in the PDS exhibits a trend of moving towards higher frequencies. They detected a 2.55~mHz break for their lowest flux observation in the hard state and observed an evolution of the inner disk radius ($R_{in}$) such that it moved towards the Innermost Stable Circular Orbit (ISCO) as the source flux increased. \\

Such low-frequency breaks in the PDS have been detected for other black hole systems as well. For \textit{e.g.}, \cite{radhika2014XTE} studied the BHXB system MAXI J1836-194 during its outburst showing the evolution of  $\nu_{brk}$ from 300-1200~mHz in its rise from low/hard state to soft state. Cygnus X-1 exhibited $\nu_{brk}$ at frequency 170~mHz in work by \cite{nowak1999cygnus} and at 200~mHz in work by \cite{belloni1996cygnus}, GRO J1655-240 showed strong band-limited noise along with break at frequencies $\sim$100–1000~mHz in \cite{mendez1998canonical}. For GX 339-4, the break frequency has been usually detected above 10~mHz, as there are very few low flux observations of this source on which detailed timing studies are performed.\\ 

In this work, we present the spectro-timing analysis of faint low/hard observations of GX 339-4 at the beginning of 2017 and 2019 outbursts. The layout of this paper is as follows: In section 2, we mention the various parameters of the source GX 339-4. In Section 3, we explain the selected AstroSat observations and reduction of data for payloads LAXPC and SXT.  In Section 4, we discuss the spectral and timing analysis. Lastly, in Section 5 we summarise and discuss the obtained results.

\begin{figure}
\caption{MAXI lightcurve for GX 339-4 in energy range 2.0-20.0~keV from June, 2009 to March, 2021 along with the zoomed in lightcurves for both AstroSat observations (Here, the vertical lines represents the AstroSat observations on $4^{th}$ and $5^{th}$ of Oct, 2017 (blue) and on $22^{nd}$ and $23^{rd}$ of Sept, 2019 (red).}
\centering
    \includegraphics[keepaspectratio, width =1.0\linewidth, height =12cm]{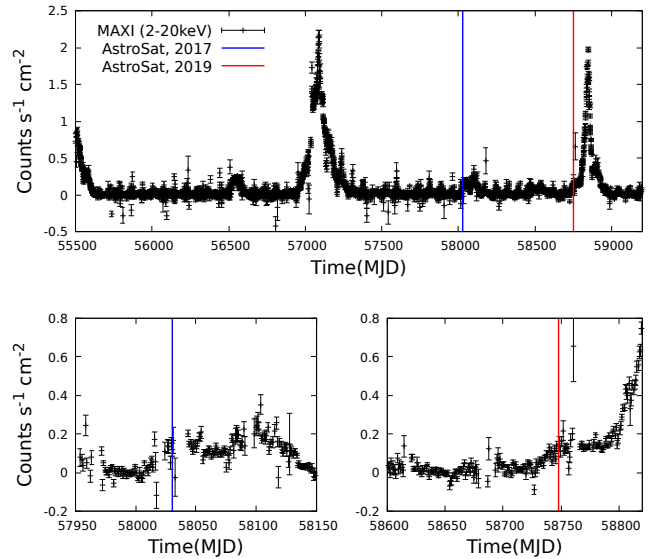}
\label{maxi_lc}      
\end{figure}

\section{Parameters of GX 339-4}
\label{sec:GX}

\cite{heida2017mass} analysed the absorption lines detected in the Near Infrared (NIR) spectrum from the donor star (K-type), when the source GX 339-4 was in its quiescent state. They estimated a mass function of f(M)~=~$1.91\pm0.08$~\(M_\odot\) by measuring the projected rotational velocity and radial velocity curve semi-amplitude of this donor star. The upper and lower limits to the mass of the black hole were set to  2.3~\(M_\odot\)~$\leq$~M~$\leq$~9.5~\(M_\odot\). \cite{hynes2004distance} estimated a distance of GX 339-4 to be 6~$\leq$~d~$\leq$~15~kpc after studying the optical spectra of Na and Ca lines along the line of sight of the source. Also, \cite{zdziarski2019} constructed the evolutionary models of the donor star of GX 339-4 and compared it with its observed data. They estimated the inclination to be between values $\sim40~\degree~-~60~\degree$. Similarly, in \cite{heida2017mass}, the upper limit of inclination was set to < 78~\degree as the binary does not show eclipses in X-ray lightcurve and the lower limit on the inclination was set to 37~\degree because of the upper limit on M~$\leq$~9.5~\(M_\odot\).

\begin{figure}
\caption{Lightcurve for (a) Source with background (b) Background (c) Source without Background for 2017 (top panel) and 2019 (bottom panel) observations.}
\centering
    \includegraphics[keepaspectratio, width = 1.0\linewidth]{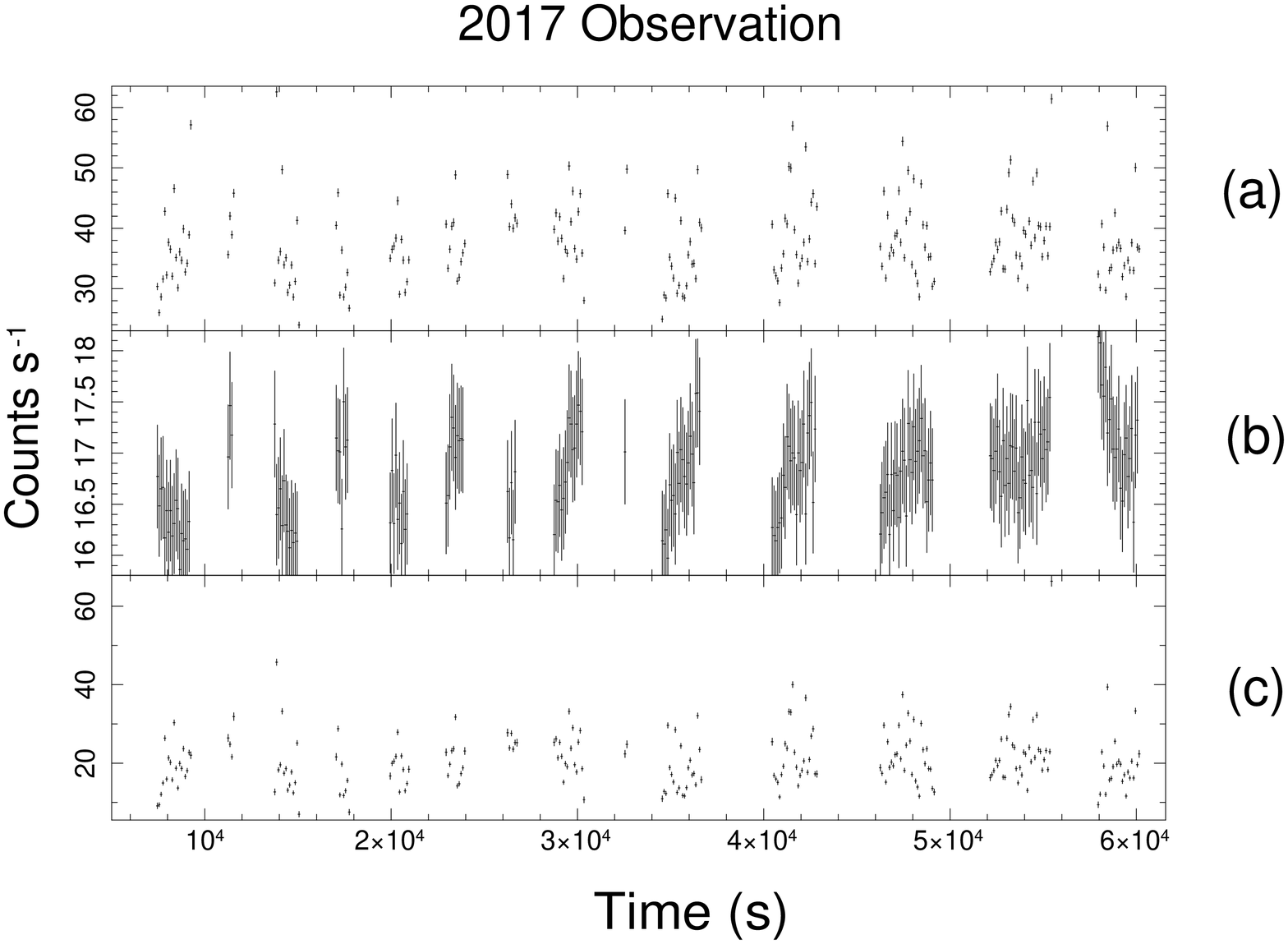}
    \vskip +2em
    \includegraphics[keepaspectratio, width = 1.0\linewidth]{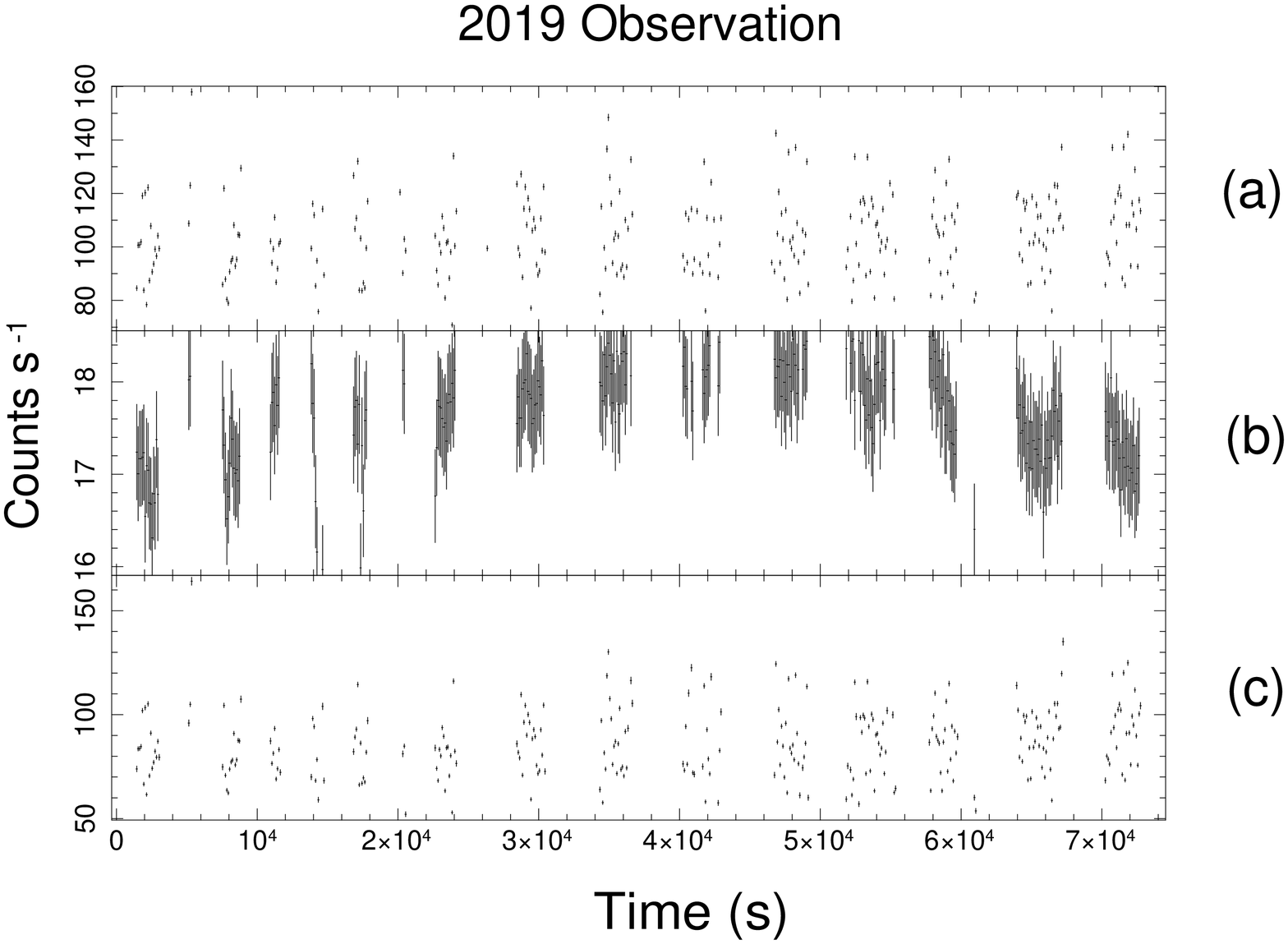}
\label{laxpc20_lc}      
\end{figure}

In this work, we have assumed the mass of the black hole to be, M, =~6~\(M_\odot\) and distance, d, =~8~kpc with inclination of the binary fixed at 60~$\degree$.\\

\section{AstroSat Observations and Data Reduction}
\label{sec:Data}

AstroSat is India's first multi-wavelength astronomical satellite, it was launched in Sept, 2015. It has five major components ascribing to different energy ranges of the electromagnetic spectrum among which Large Area X-ray Proportional Counter (LAXPC) and Soft X-ray Telescope (SXT) provide opportunity to study X-ray emission in energy ranges 3.0-80.0~keV and 0.3-8.0~keV respectively \citep{yadav2016astro, agrawal2017astro}. \\

The source GX 339-4 exhibits outburst every 2-3 years amongst which some outbursts fail to show state transitions and remain in the low/hard state. The source was in the course of one such failed outburst in September 2017, when after staying in quiescence  since 2016, the source re-brightened and was observed by Faulkes telescope \citep{Faulkes}. Follow up observations were conducted by XRT/SWIFT \citep{Gandhi} and Australia Telescope Compact Array (ATCA) \citep{RussellRadio}. Later on, AstroSat observed it at the initial stage of this outburst on $4^{th}$ and $5^{th}$ of Oct, 2017 (Observation ID: A04\_109T01\_9000001578) with an exposure time of $\sim$60~ks. The source again re-brightened in August 2019 \citep{AtelAnjaliRao} following which AstroSat observed it on $22^{nd}$ and $23^{th}$ of Sept, 2019 (Observation ID: A05\_166T01\_9000003192) with an exposure time of $\sim$35~ks. This observation was also extracted at the beginning of the outburst. The MAXI light curve for the source in energy range 2.0-20.0~keV is shown in Figure \ref{maxi_lc} with simultaneous AstroSat observations represented as vertical lines. We did not consider early AstroSat observations of the source in this work due to their significantly lower values of LAXPC count rates of $<$ 3 counts s$^{-1}$ as compared to 22 and 86 counts s$^{-1}$ for 2017 and 2019 observation respectively. Here, in Figure \ref{laxpc20_lc}, we plot the total count rate of source with background, background and background subtracted source count rate using LAXPC20 data for both observations in energy range 3.0-20.0~keV. We have analysed the data from LAXPCs and SXT and the reduction of their data-sets is discussed in the next section. 

\begin{figure*}
      
    \caption{The joint photon spectrum with ratio of data to model plot for GX 339-4 in energy range 0.6-20.0 ~keV for 2017 observation (upper panel) with models ($A$) \textit{TBabs*Powerlaw} ($B$) \textit{TBabs*Ireflect*Powerlaw} ($C$) \textit{TBabs*(Powerlaw+Relline)} and for 2019 observation (bottom panel) with models ($A'$) \textit{TBabs*Powerlaw} ($B'$) \textit{TBabs*Ireflect*Powerlaw} ($C'$) \textit{TBabs*(Powerlaw+Relline)}.}
    
   \hspace{-6cm}
    \includegraphics[width = 6.3cm]{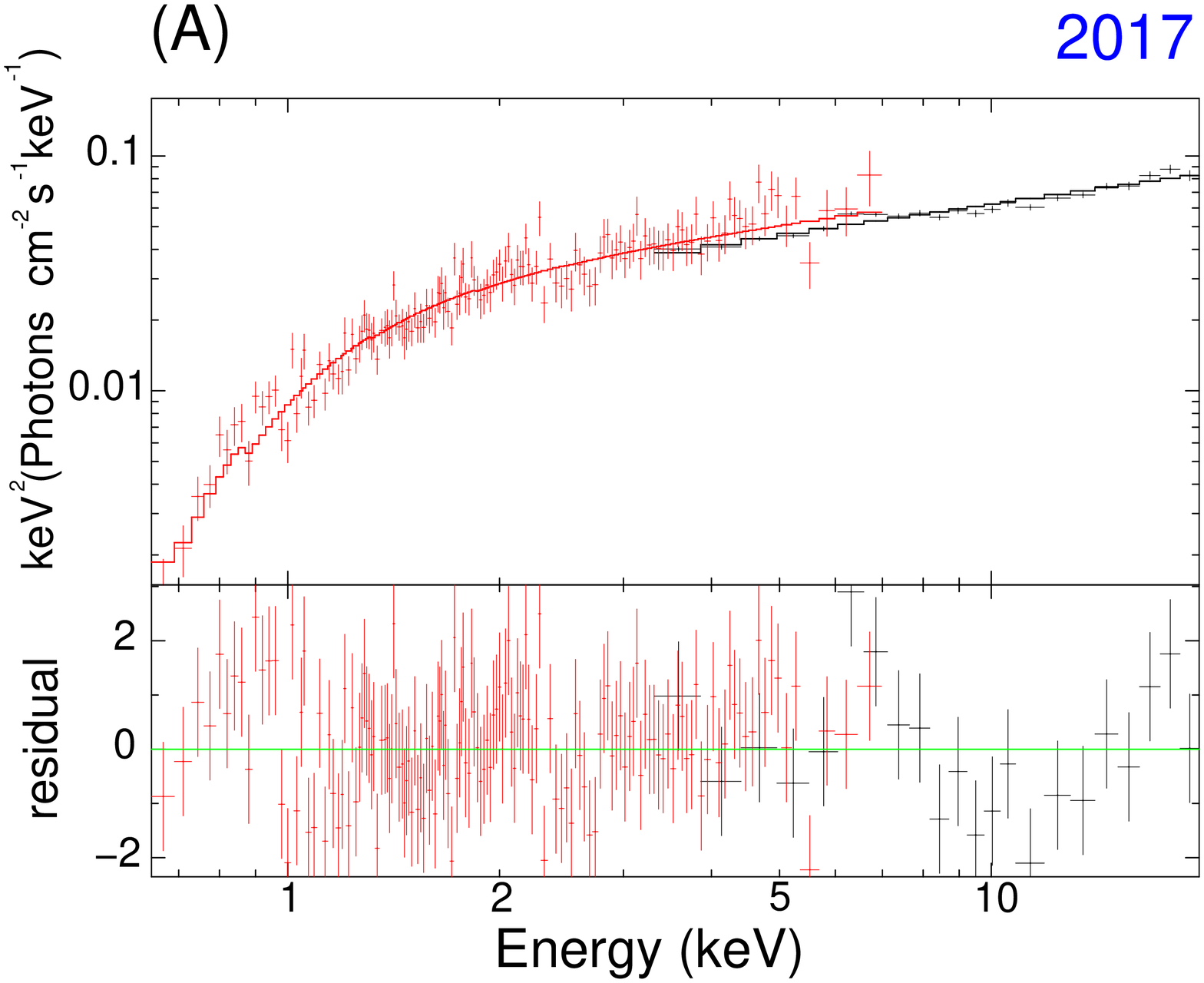}
    \includegraphics[width = 6.3cm]{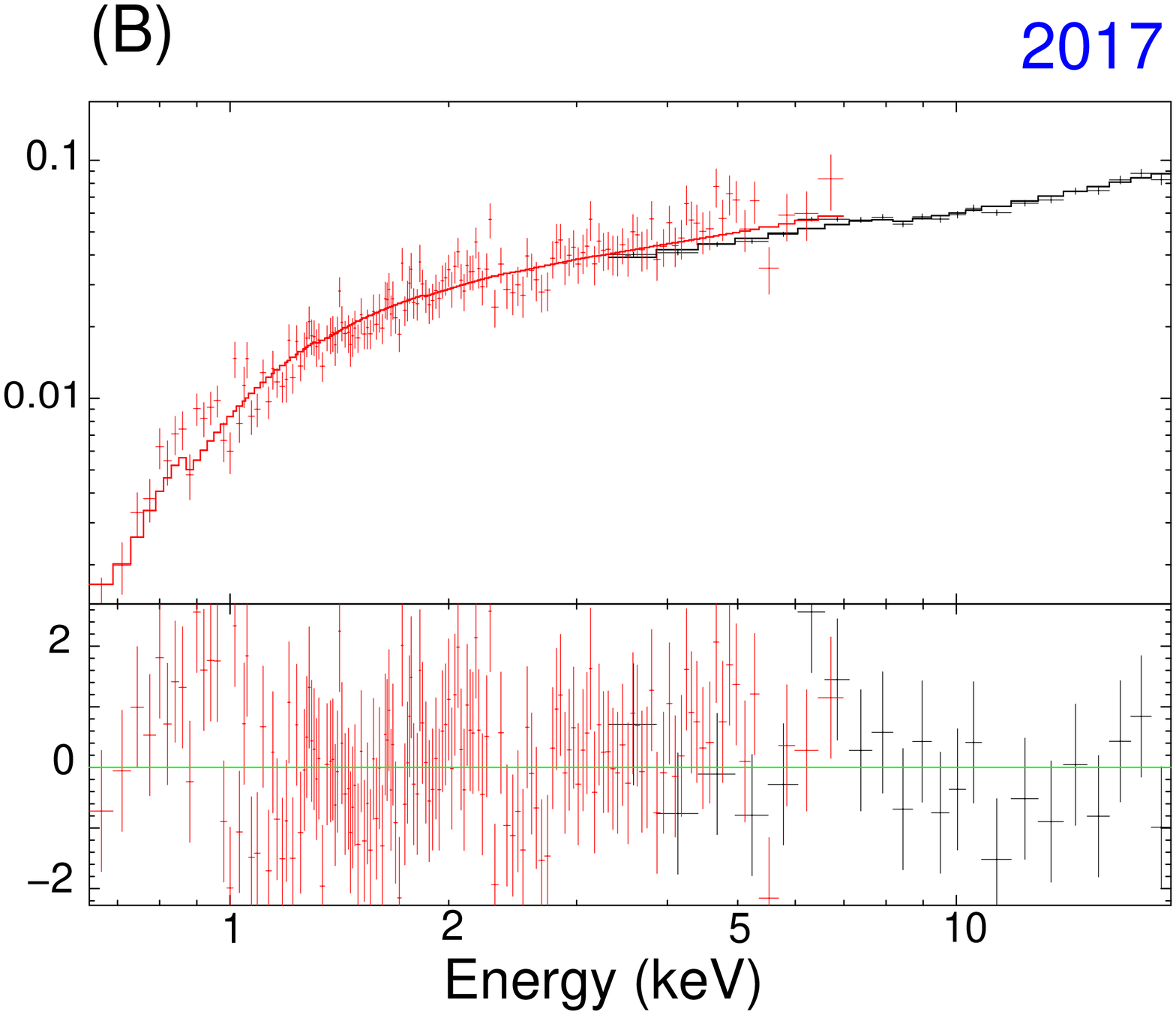}
    \includegraphics[width = 6.3cm]{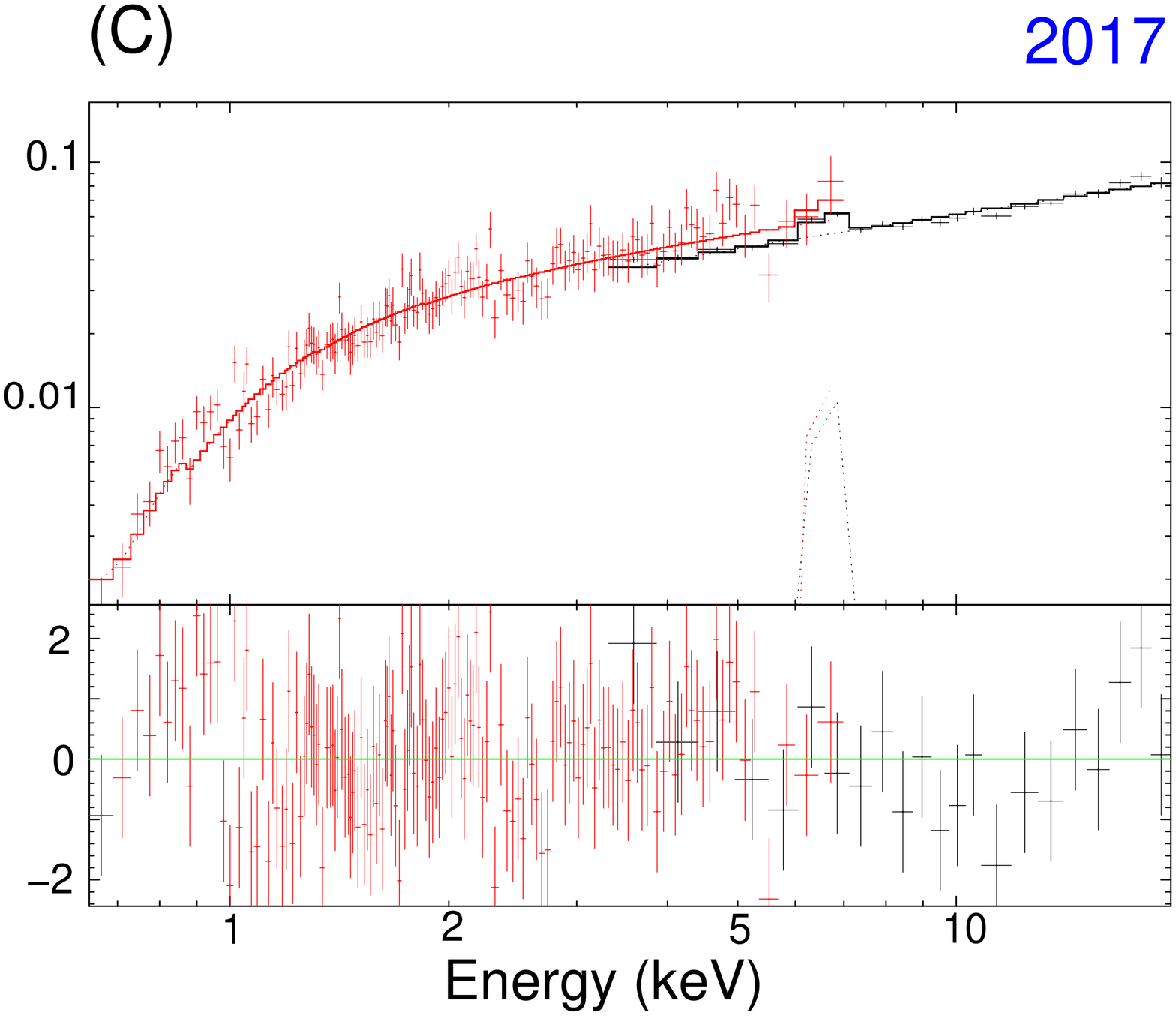}
  \hspace{-6cm}
  
   \label{fig:Spectral}
  
\end{figure*}
  
\begin{figure*}

    \hspace{-6cm}
    \includegraphics[width = 6.3cm]{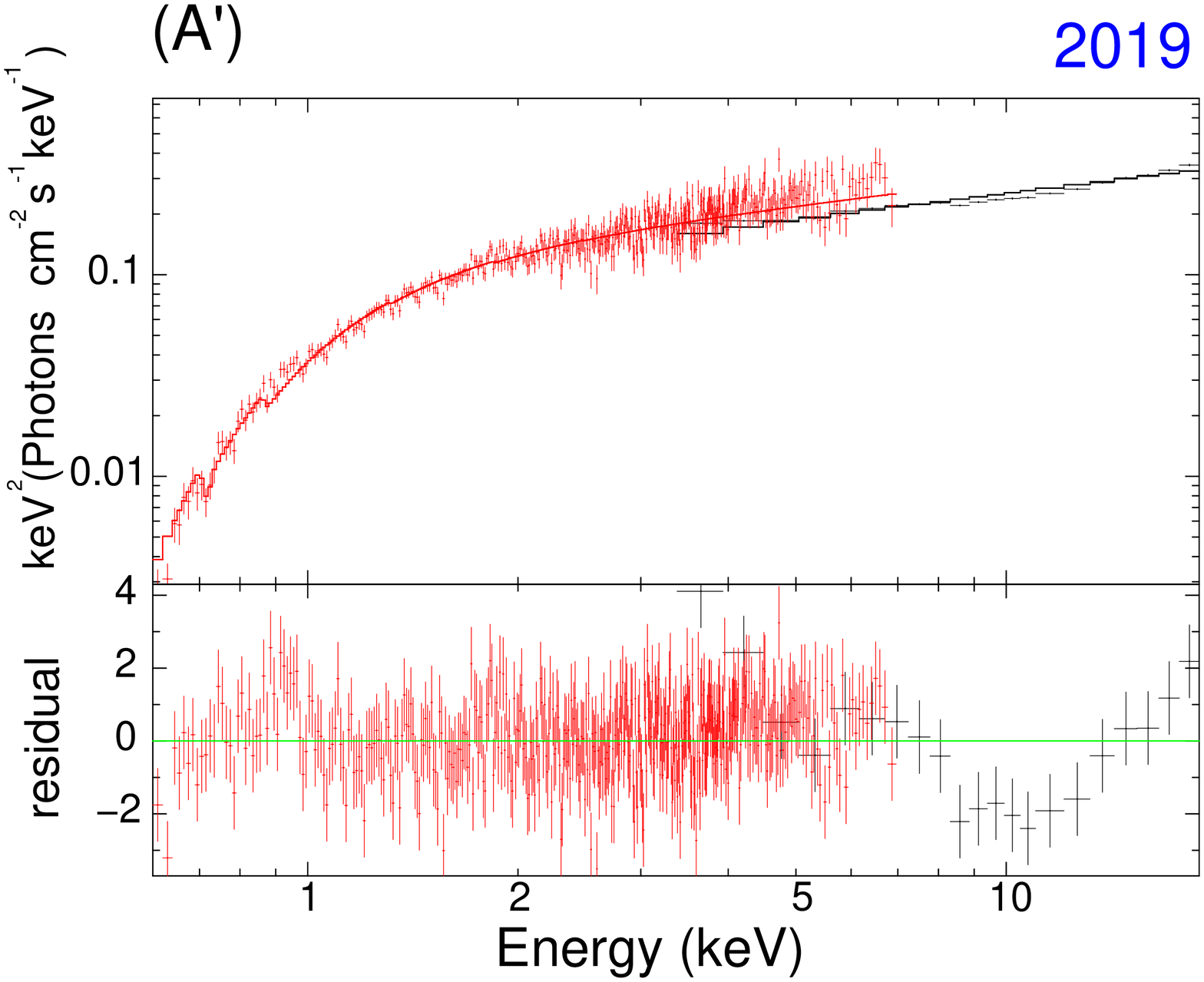}
    \includegraphics[width = 6.3cm]{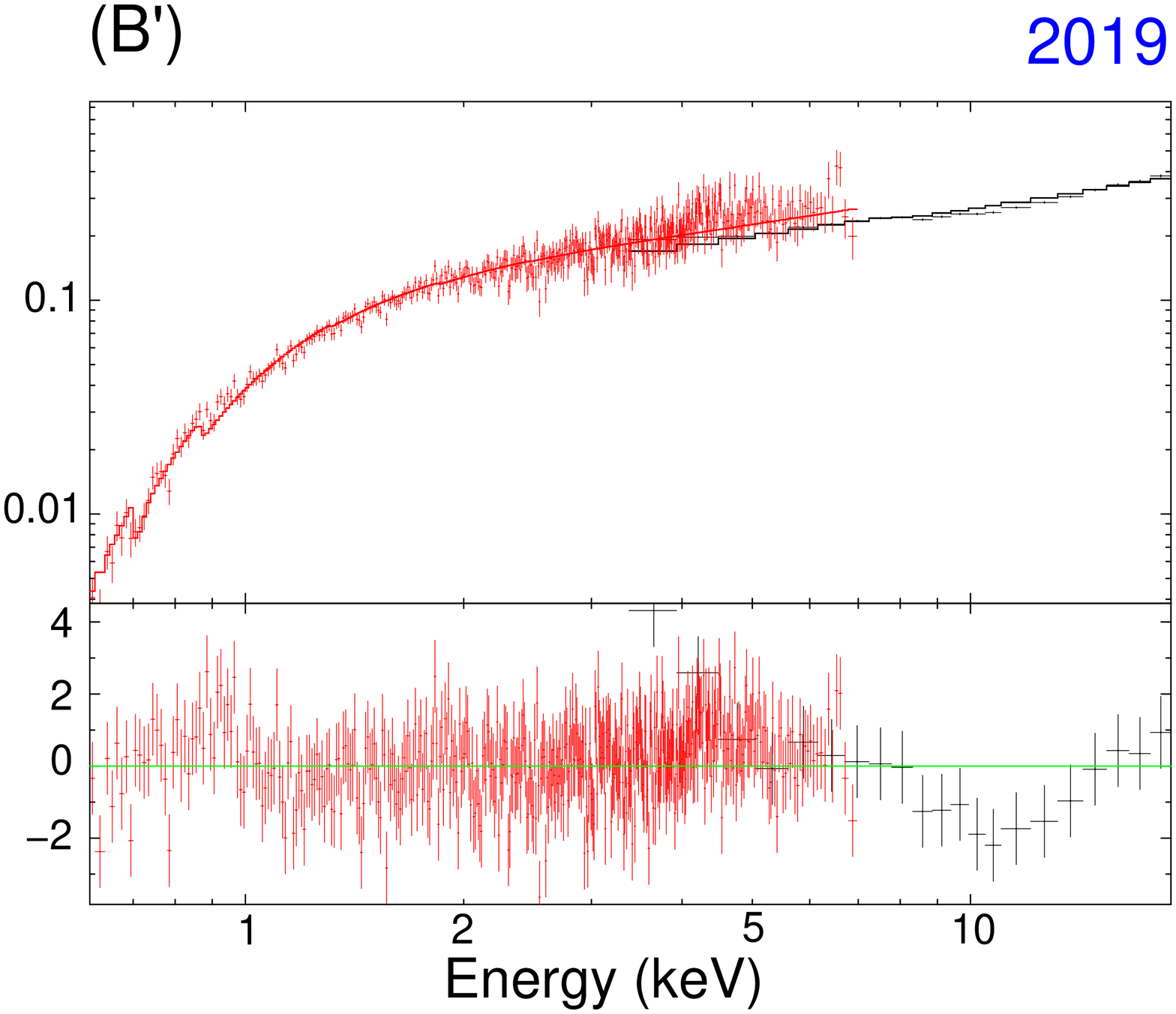}
        \hskip +3.0ex
    \includegraphics[width = 6.0cm]{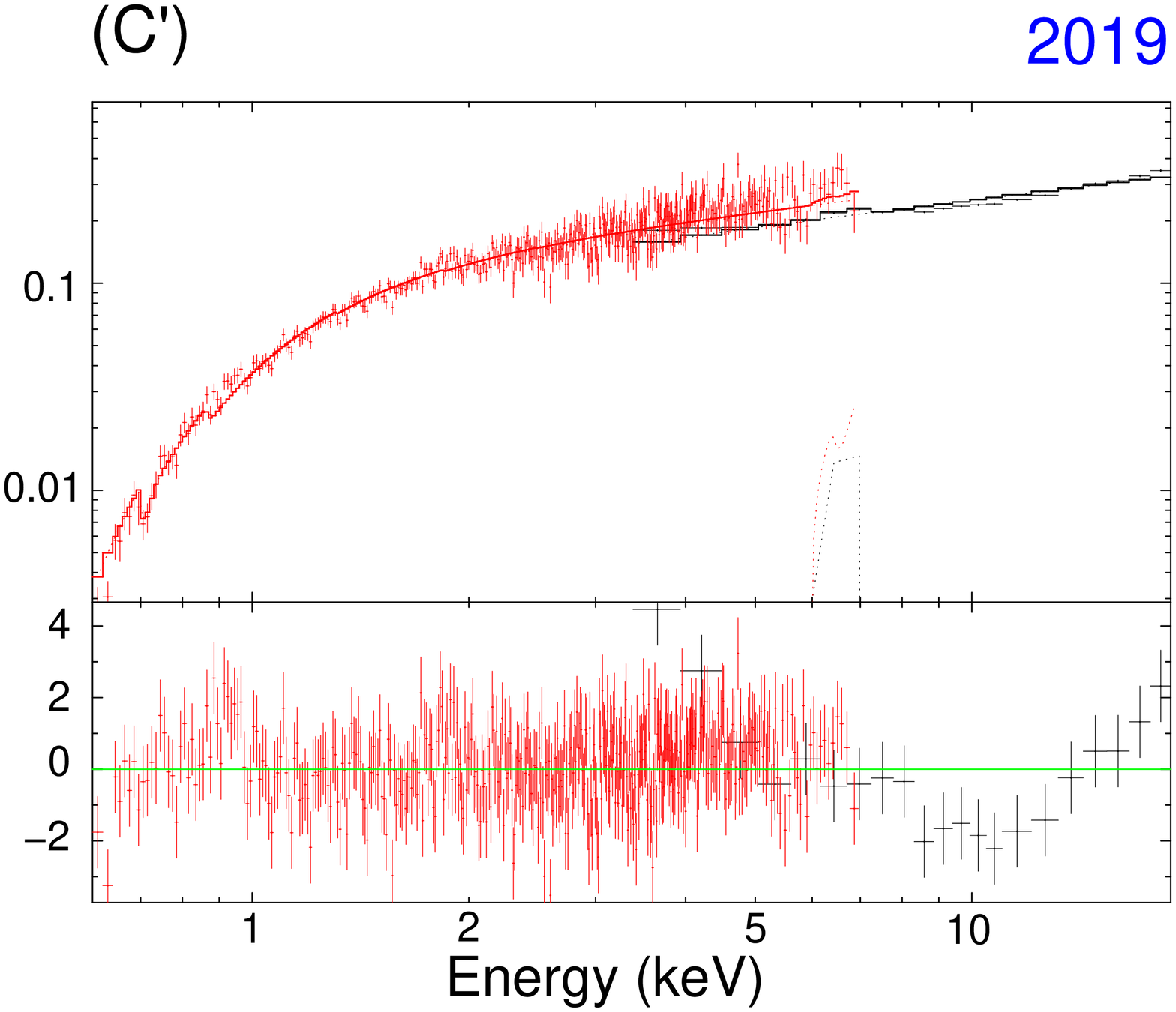}
   \hspace{-6cm}
   
\end{figure*}

\renewcommand\arraystretch{1.5}
\begin{table*}
\centering
\caption{Best fit values of spectral parameters for both observations of 2017 and 2019 with models  \textit{TBabs*Powerlaw} (Top Panel), \textit{TBabs*Ireflect*Powerlaw} (Middle Panel) and \textit{TBabs*(Powerlaw+Relline)} (Bottom Panel) along with the Eddington ratio for each model.}

\begin{tabular}{c|cccc} \hline \hline

\textbf{Model} & \textbf{Parameter} & \textbf{2017} &  \textbf{2019}\\

\hline \hline

\textsc{{TBabs}} & $N_H$ ($10^{22}$ $cm^{-2}$) & $0.430^{+0.035}_{-0.033}$ & $0.440^{+0.014}_{-0.014}$ \\

\textsc{{Powerlaw}} & $\Gamma$ & $1.570^{+0.031}_{-0.031}$ & $1.583^{+0.020}_{-0.020}$\\
                          
                           & $N_{\Gamma}~(10^{-2})$ & $2.35^{+0.17}_{-0.15}$ & $9.73^{+0.45}_{-0.43}$  \\
                           
\textsc{{$\chi^2/dof$}}  & & 189.61/169.00 = 1.12 &  531.82/441.00 = 1.21\\

\textsc{{$L/L_{edd}~\%$}}  & & 0.114 &  0.463 \\

\hline

\textsc{{TBabs}} & $N_H$~($10^{22}$ $cm^{-2}$) & $0.490^{+0.050}_{-0.047}$ & $0.460^{+0.022}_{-0.021}$ \\

\textsc{{Powerlaw}} & $\Gamma$ & $1.680^{+0.080}_{-0.077}$ & $1.640^{+0.043}_{-0.042}$ \\
                           & $N_{\Gamma}~(10^{-2})$ & $2.68^{+0.30}_{-0.27}$ & $10.30^{+0.66}_{-0.62}$ \\

\textsc{{Ireflect}} & RS & $0.50^{+0.45}_{-0.35}$ & $0.27^{+0.22}_{-0.19}$ \\

\textsc{{$\chi^2/dof$}}  &  & 183.55/168.00 = 1.09 &  524.99/440.00 = 1.19\\
 
\textsc{{$L/L_{edd}~\%$}}  & & 0.115 &  0.464 \\      

\hline

\textsc{{TBabs}} & $N_H$ ($10^{22}$ $cm^{-2}$) & $0.410^{+0.034}_{-0.033}$ & $0.440^{+0.014}_{-0.014}$\\

\textsc{{Powerlaw}} & $\Gamma$ & $1.550^{+0.032}_{-0.032}$ & $1.58^{+0.02}_{-0.02}$\\
                           & $N_{\Gamma}$ ($10^{-2}$) & $2.31^{+0.17}_{-0.15}$ &$9.59^{+0.46}_{-0.44}$ \\

\textsc{{Relline}} & $N_{Rell} $ ($10^{-4}$) & $2.31^{+1.06}_{-1.03}$ & $3.77^{+3.2}_{-3.1}$ \\

\textsc{{$\chi^2/dof$}}  &  & 175.60/168.00 = 1.05 &  522.67/440.00 = 1.19 \\

\textsc{{$L/L_{edd}~\%$}}  & & 0.110 &  0.460 \\

\hline \hline

\end{tabular}
\\
\begin{flushleft}
\hspace{2cm}Notes :  
1) The errors were calculated in 90~\% confidence region. \\
\hspace{2.8cm}  2) RS is reflection scaling factor. \\
\hspace{2.8cm}  3) Eddington ratio ($L/L_{edd}~\% $) was calculated using \textit{cflux} model in energy range 0.6-10.0~keV.
\end{flushleft}
\label{Spectral_fits}
\end{table*}

\subsection{Large Area X-ray Proportional Counter(LAXPC)}
\label{sec:LAXPC}

LAXPC has three identical proportional counters LAXPC10, 20 and 30. LAXPC30 was switched off earlier on $8^{th}$ March, 2018 because of abnormal gain changes \url{http://astrosat-ssc.iucaa.in/}. LAXPC provides a large effective area for energies 3.0-80.0~keV ($\sim6000$~$cm^2$) and high time resolution of 10~$\mu$s \citep{agrawal2017astro}. Level 1 data for both 2017 and 2019 observations were downloaded from AstroSat archive. These observations were conducted in Event mode which records data with information about the arrival time and energy of each incoming photon. Reduction of level 1 raw data to level 2 data was completed with the LAXPC software (version as of May, 2018) available on \url{http://astrosat-ssc.iucaa.in/?q=laxpcData}.
Due to the low count rate of both observations, see Figure \ref{laxpc20_lc},  the background lightcurves and spectra were generated using the Fortran codes for Faint sources in LAXPC software. The lightcurve and spectra were obtained using "\textit{laxpc\_make\_lightcurve}" and "\textit{laxpc\_make\_spectra}" codes. \par

\subsection{Soft X-ray Telescope(SXT)}
\label{sec:SXT}

SXT observes low energy or softer X-rays in energy range 0.3-8.0~keV with high sensitivity. This payload has a Wolter I optics geometry arrangement of reflecting mirrors and an X-ray CCD, which provides an effective area of approximately 90~$cm^2$ at $\sim1.5~keV$ \citep{singh2017soft}. The level 2 data for SXT was downloaded from AstroSat archive in Photon Counting mode (PC). The reduction of data was achieved with tools provided by SXT team at TIFR (\url{https://www.tifr.res.in/~astrosat_sxt/dataanalysis.html}). All individual clean event files  for all orbits were merged with SXT\_EVENT\_MERGER tool for each data set. The image of the source was extracted using the merged eventfile and a circular region of radius 15~arcmin centered on the source was considered for further data analysis. Also, latest RMF (sxt\_pc\_mat\_g0to12.rmf) and Background (SkyBkg\_comb\_EL3p5\_Cl\_Rd16p0\_v01.pha) files were used. The ARF file (sxt\_pc\_excl00\_v04\_20190608.arf) given by the SXT team was corrected for the vignetting effect using the SXT\_ARF\_TOOL.  The spectrum for SXT was extracted using the HEASoft (version 6.26.1) package XSELECT. The SXT spectrum was further grouped with the background, RMF and corrected ARF files by the interactive command 'grppha' producing the final SXT spectrum.  \par

\section{Analysis and Results}
\label{sec:Analysis}

\subsection{Spectral Analysis}
\label{sec:Spectral}

The spectrum of most black hole binaries has been observed to have three major components with varying contributions during different spectral states (Hard or soft states), first is multi-color black body radiation emitted from the surface of the thin optically thick accretion disk, second is comptonization spectrum from the inner hot flow where the seed photons from the disk get Compton scattered by the high energy electrons in the Compton cloud giving the hard spectrum and lastly the reflection spectrum, where the contribution comes from the irradiation of accretion disk by these comptonized photons giving rise to absorption and emission features such as Fe k~$\alpha$ line in range 6.4-6.7~keV and Compton hump peaking at $\sim$30~keV. Since, the skewness and broadening of such reflection features seen in the spectrum are induced by the relativistic effects occurring in the vicinity of the compact object, therefore these features in turn allow probing the region of strong gravity. \\

For spectral analysis, we considered joint fitting of data from LAXPC20 and SXT, this fitting allows the same model to fit two data files simultaneously. Only LAXPC20 data was used for spectral fitting because of its better constraint on the background. For each observation, the common usergti was created for payloads LAXPC20 and SXT. This usergti was then used to create the spectrum for LAXPC20 in energy range 3.0-20.0~keV (for >20.0~keV the background dominates the spectrum notably) and the spectrum of SXT in energy range 0.6-7.0~keV.  The parameters for all the selected models were tied for data from the payloads LAXPC20 and SXT. A constant factor was used to account for the two different detectors. Also, for all fittings a 3$\%$ of systematic error was considered.  \\

In the low/hard state, the continuum is dominated by comptonized spectrum, therefore the  first model we fit to the photon spectrum for both observations is an absorbed power-law (XSPEC model \textit{TBabs*Powerlaw}) to the joint spectrum of LAXPC20 and SXT in total energy range of 0.6-20~keV. Here, the model component \textit{TBabs} accounts for the intergalactic absorption of the incoming photons from the source, $N_H$ gives the column density in units of {$10^{22}~cm^{-2}$}. We let this parameter vary in the fit. The fitted model parameters along with Reduced $\chi^2$ = $\chi^2/dof$, where dof is the degrees of freedom, are given in Table \ref{Spectral_fits}. The fit shows that the continuum for both observations could be described by the absorbed power-law model. The photon index takes value <~2 for both observations ($\sim$1.57 for 2017 and $\sim$1.58 for 2019) which implies that during these observations the source was in the hard state, which is expected for this source as these observations were extracted at the beginning of the two outbursts.

After fitting the spectra with absorbed power-law, some residuals still remain in the data to model ratio. Therefore, the  next model we fit the joint spectrum is \textit{Ireflect} convolved with \textit{Powerlaw} (XSPEC model \textit{TBabs*Ireflect*Powerlaw}). Reflection model \textit{Ireflect} is a generalisation of model \textit{Pexriv}, XSPEC \citep{magdziarzpexriv}. This model accounts for the edge absorptions and ignores the line emissions such as the Fe line. In this model combination, the abundances of Fe and other elements were set to solar values. We also assume the temperature of accretion disk to be 30,000~K and since the ionization parameter was not well constrained therefore we fix it to a large value of 900. The best fit values of parameters for both observations are given in Table \ref{Spectral_fits}. \\

Applying the above model to the spectral modelling improved the fit with $\Delta\chi^2\sim6$ for 2017 observation and $\Delta\chi^2\sim7$ for 2019 observation but residuals in the iron line region persist therefore we use another model combination consisting of reflection component \textit{Relline} \citep{DauserRelline} (XSPEC model  \textit{TBabs*(Powerlaw+Relline)}). One of the most prominent features seen in the reflection spectrum of black hole binary system is Fe k$\alpha$ line in region 6.4-6.7~keV (depending upon the ionization of Fe in the disk). \textit{Relline} applies appropriate relativistic broadening effects to the Fe k$\alpha$ line after assuming an intrinsic zero width for the line emission.  We fit the model \textit{TBabs*(Powerlaw+Relline)} to the joint spectrum of LAXPC20 and SXT. The emissivity indices for both inner and outer disk were set to 3 (Emissivity depends on radius, R, as $\epsilon$ $\propto$ $R^{-3}$) which is an assumed value for Newtonian accretion disks \citep{fabian1989emis,laor1991line}. We fix the outer radius of disk to  1000~$R_g$, where $R_g$ is the gravitational radius~($GM/c^2$). Since the fit was insensitive to the value of spin therefore we fixed it to value 0.9. $R_{in}$ could not be constrained and the $\chi^2$ fit was insensitive to its different values, therefore, we fixed it to a fiducial value of 100~$R_g$. After including this reflection component in the model the $\Delta\chi^2$ improved further by $\sim8$ for 2017 observation and by $\sim2$ for 2019 observation.  The respective fit parameters are mentioned in Table \ref{Spectral_fits}. There are residuals in the LAXPC spectrum for the 2019 observation, which could either be due to a slight underestimation of the systematic uncertainty in the response (3$\%$ has been assumed here) or a physical component. We note that the addition of an absorption edge at $8.5$ keV (perhaps due to iron) and with a depth of $\sim 0.23$ significantly reduces the residuals and improves the fit, $\Delta \chi^2 \sim 45$. However,  given the uncertainty in the LAXPC response, we are not sure about the physical origin of the edge. We note that the addition of the component does not lead to any significant change in the value of the power-law index or the flux level. Furthermore, we also fit two other commonly used reflection components \textit{Xillver} and \textit{Reflionx} and we do not mention the best fit values as the $\Delta \chi^2$ did not improve further for both observations. \\

\begin{figure*}
      \caption{Fitting of $\nu$~$P_\nu$ with multi-Lorentzian model for (A) LAXPC10 (B) LAXPC20 (C) LAXPC10 + LAXPC20 (D) SXT for 2017 observation.}
   \centering
    
    \includegraphics[height = 6.5cm, width = 8.5cm]{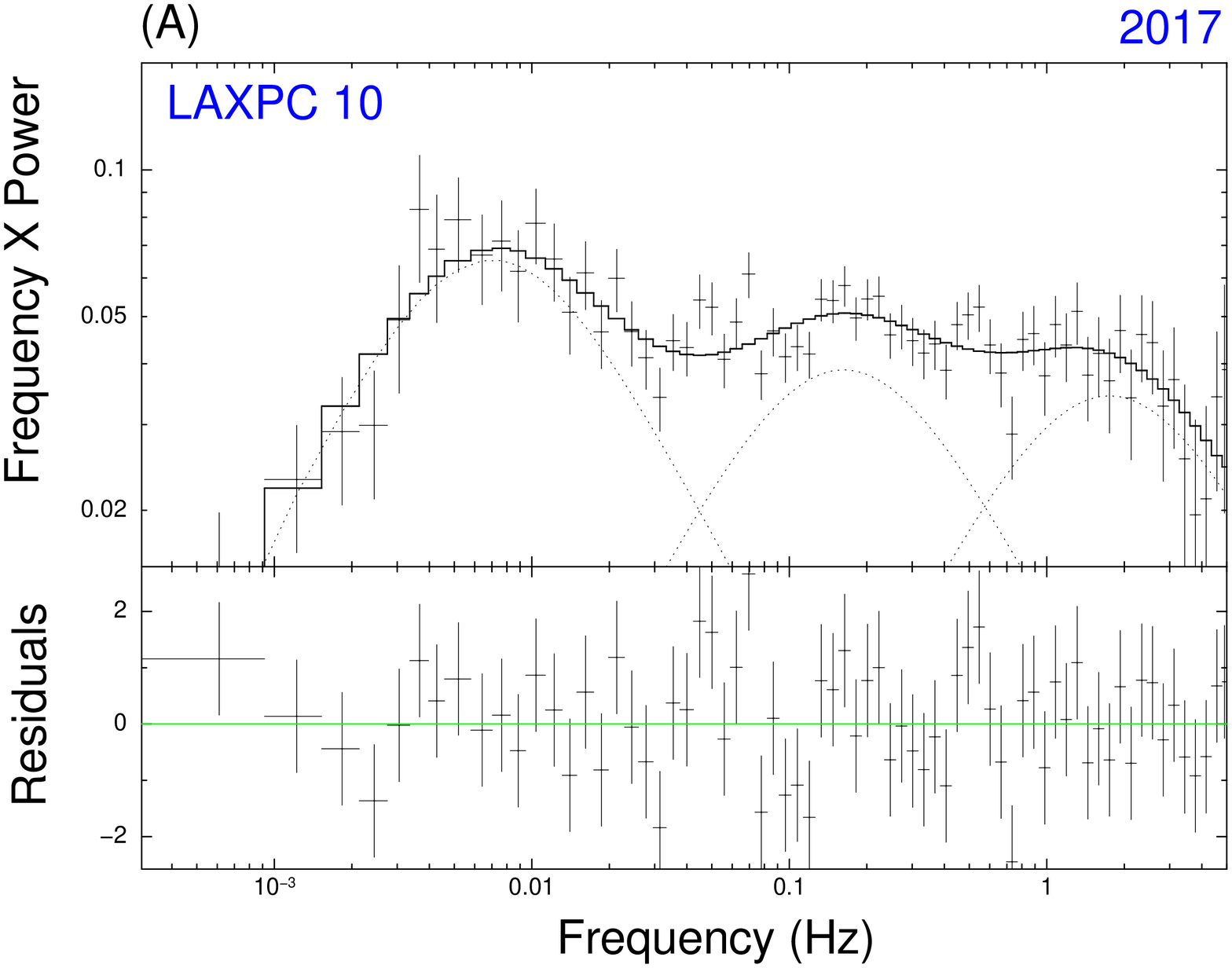}
    \hskip +6.0ex
    \includegraphics[height = 6.5cm, width = 8.5cm]{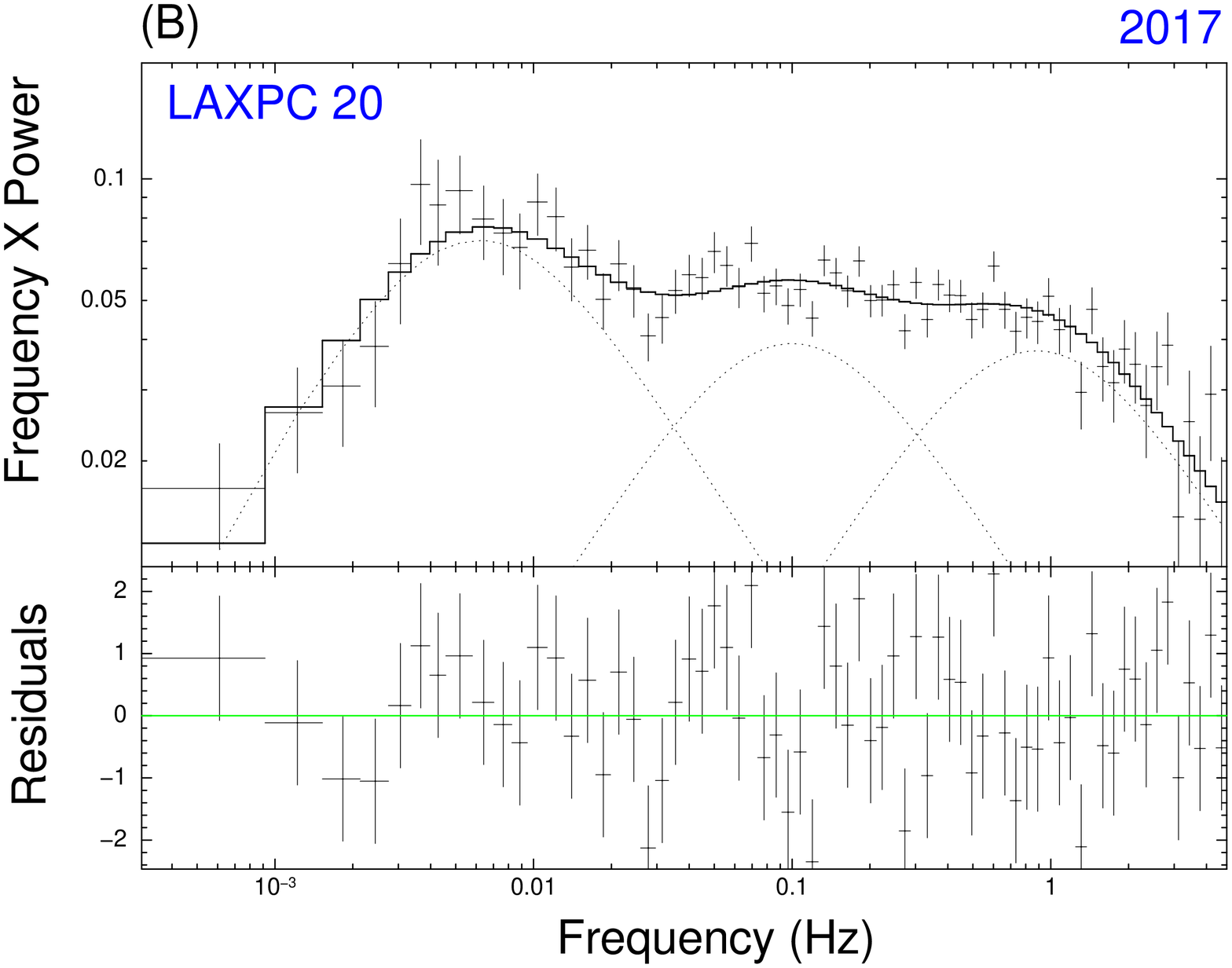}
    \includegraphics[height = 6.5cm, width = 8.5cm]{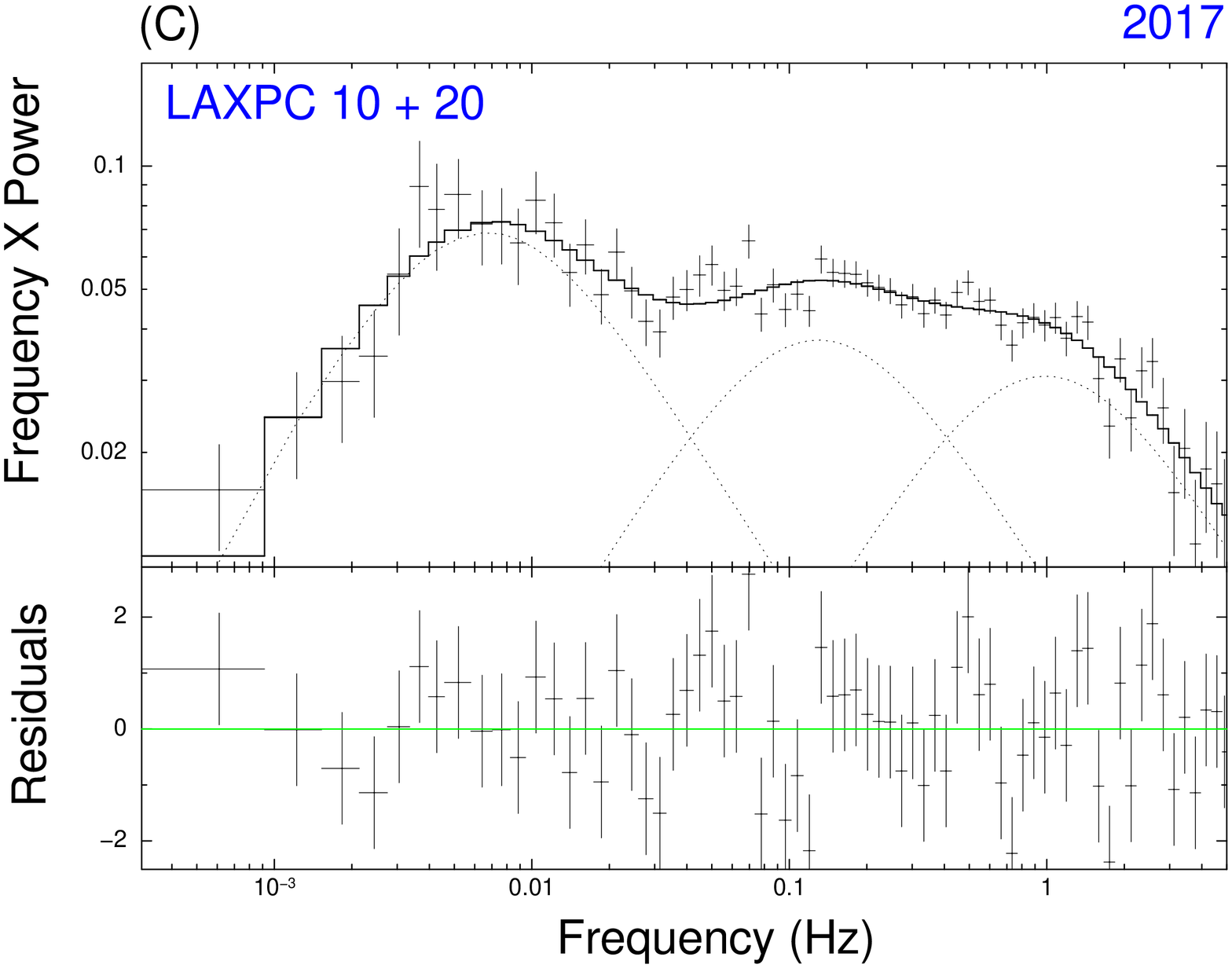}
    \hskip +6.0ex
    \includegraphics[height = 6.5cm, width = 8.5cm]{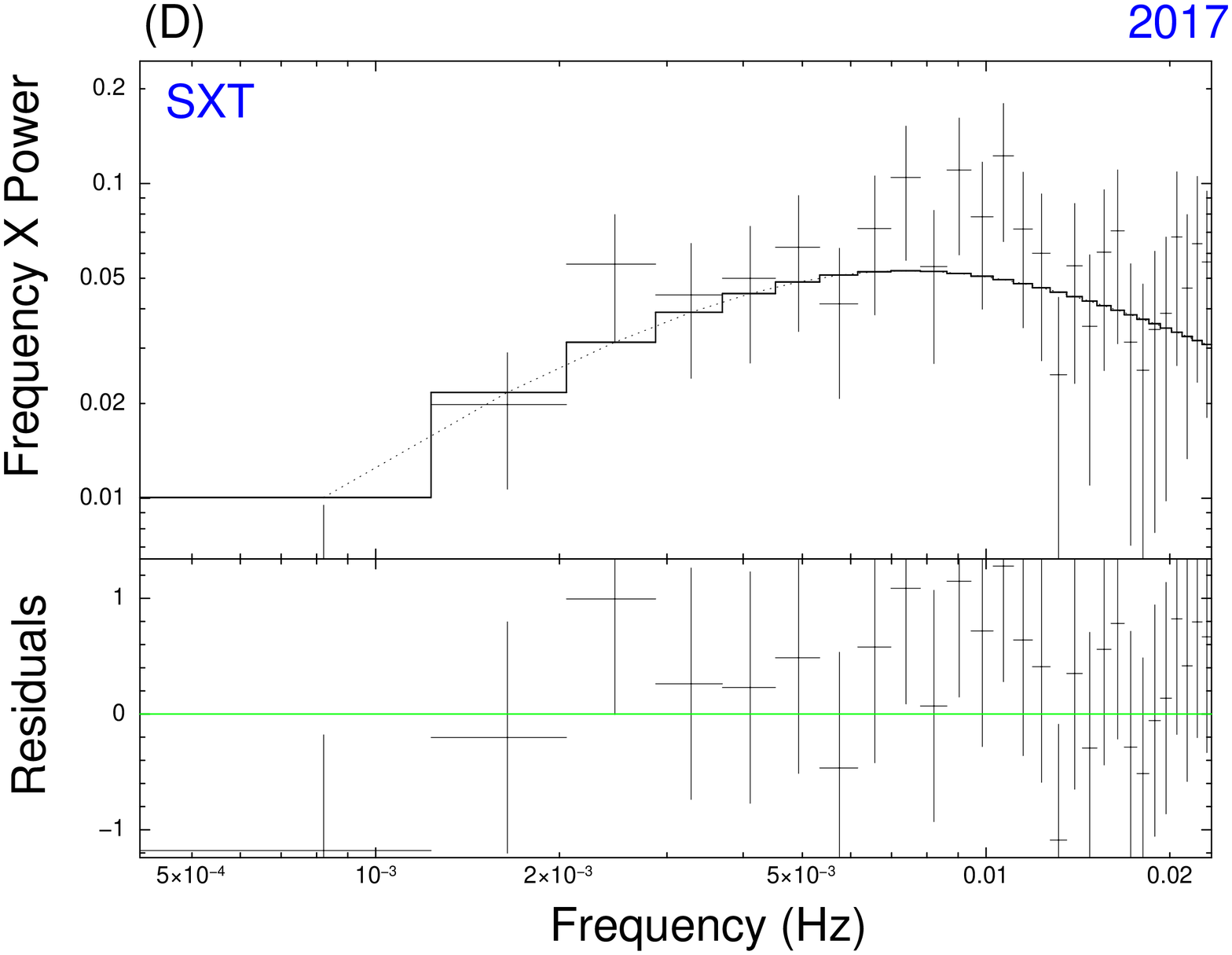}
    
   \label{fig:PDS_lorentz_2017}

\end{figure*}

\begin{figure*}
   \vspace{1cm} \caption{Fitting of $\nu$~$P_\nu$ with multi-Lorentzian model for (A) LAXPC20 and (B) SXT for 2019 observation.}
   \centering
    \includegraphics[height = 6.5cm, width = 8.5cm]{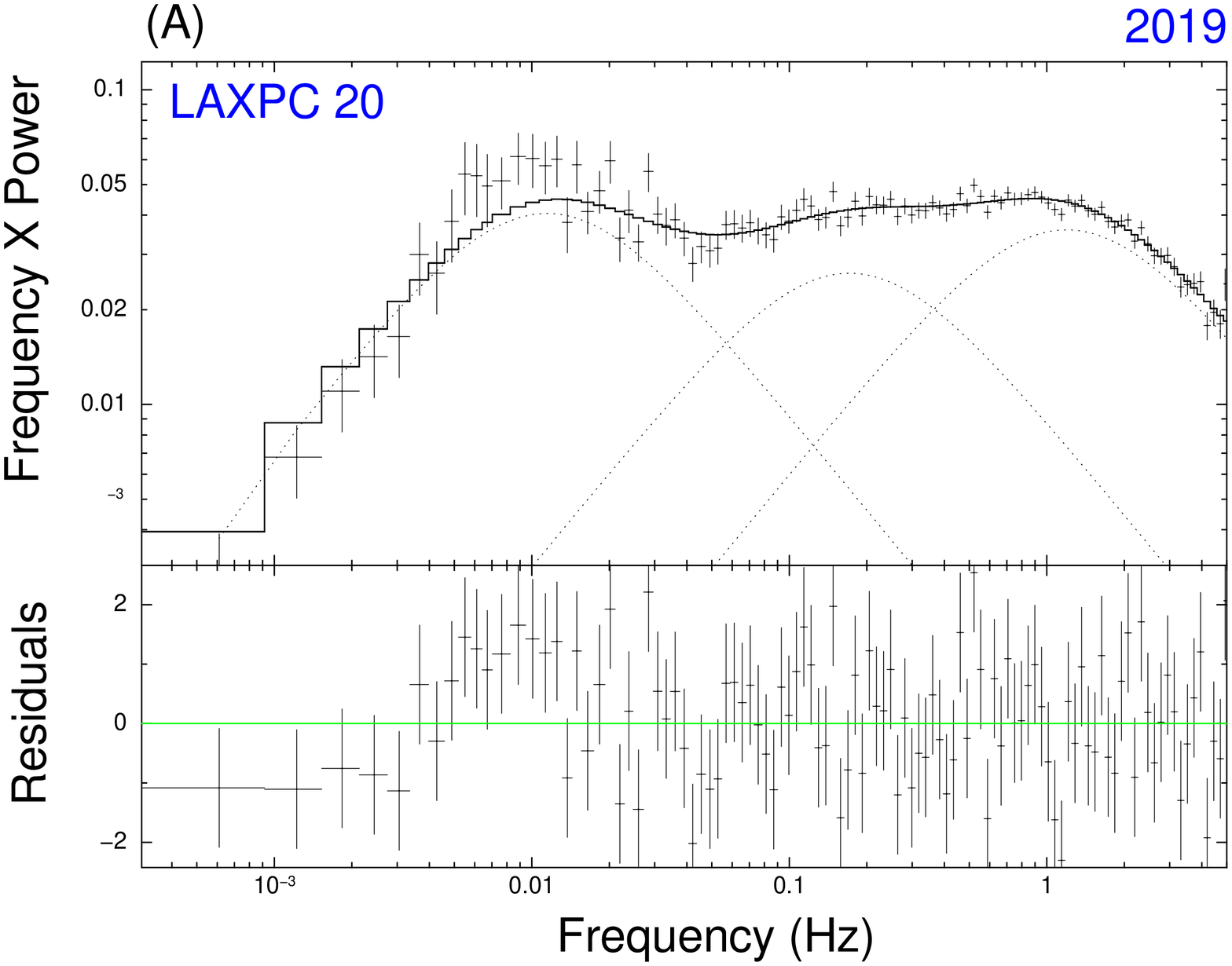}
    \hskip +6.0ex
    \includegraphics[height = 6.5cm, width = 8.5cm]{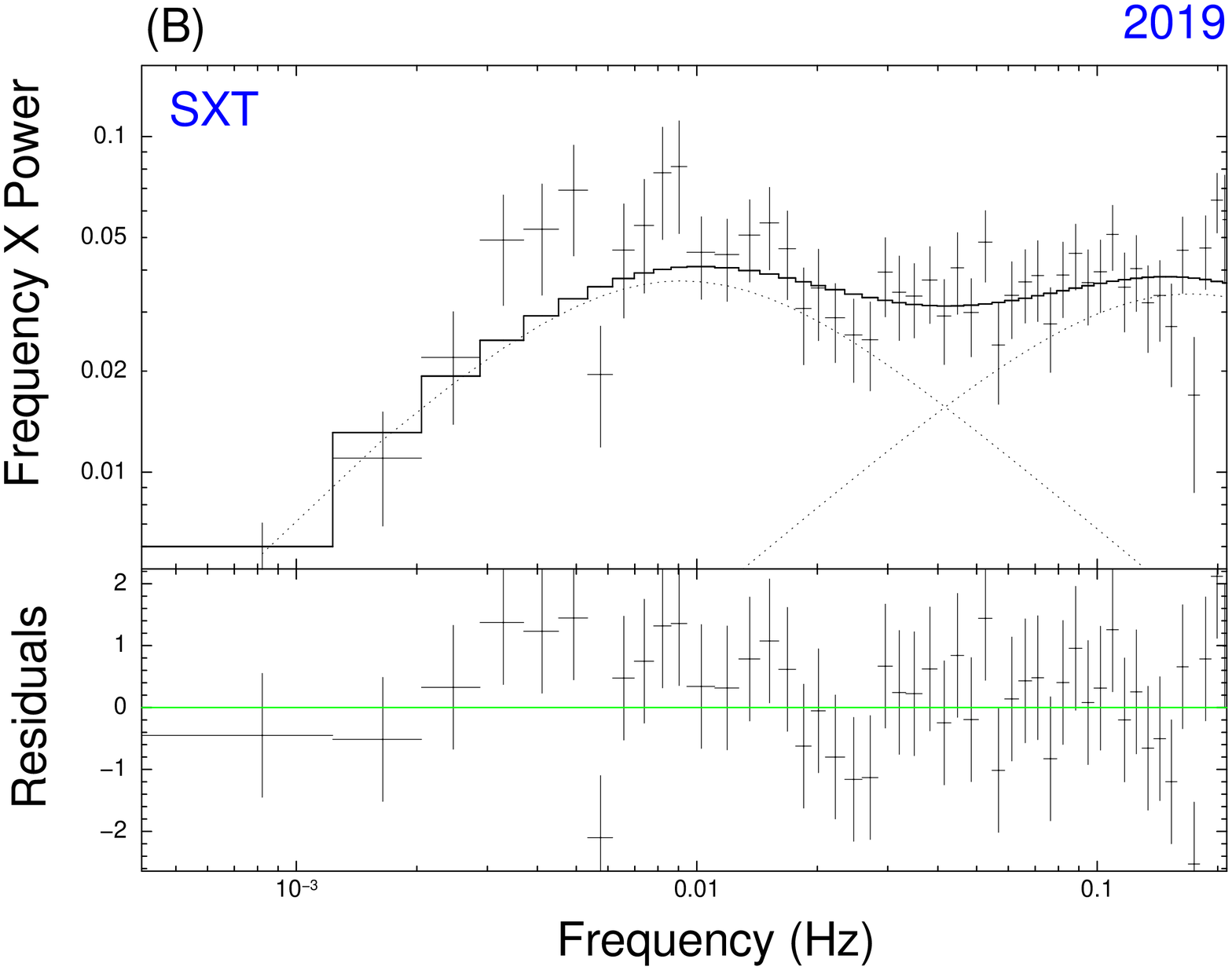}

   \label{fig:PDS_lorentz_2019}

\end{figure*}

\subsection{Timing Analysis}
\label{sec:Timing}

\renewcommand\arraystretch{2.0}
\begin{table*}
\centering
\caption{Best fit parameters estimated with multi-Lorentzian fitting of PDS for both observations.}

\begin{tabular}{p{1.5cm}p{1.2cm}|p{0.8cm}p{1.2cm}p{0.8cm}p{1.0cm}p{0.8cm}|p{0.8cm}p{1.0cm}|p{2.2cm}}

\hline
Observation & Payload & $\nu_{L1}~(mHz)$ &   $N_{L1}$ & $\nu_{L2}~(Hz)$  & $N_{L2}$ & $\nu_{L3}~(Hz)$  & $N_{L3}$ &  $N_{noise}$ & $\chi^2$/dof \\

\hline \hline

& {\textsc{LXP10}} & $6.95^{+1.11}_{-0.93}$ & $0.21^{+0.02}_{-0.02}$ & $0.16^{+0.03}_{-0.03}$ & $0.12^{+0.01}_{-0.01}$ &  $1.75^{+1.16}_{-0.64}$ & $0.11^{+0.06}_{-0.028}$ & $0.232^{+0.008}_{-0.008}$ & 63.48/63.00=1.01\\
\textbf{2017} & {\textsc{LXP20}} & $6.12^{+1.10}_{-0.93}$ & $0.22^{+0.03}_{-0.03}$  & $0.196^{+0.054}_{-0.043}$ & $0.12^{+0.01}_{-0.01}$ & $0.86^{+0.29}_{-0.19}$  & $0.118^{+0.013}_{-0.014}$ & $0.182^{+0.002}_{-0.003}$ & 83.67/63.00=1.33\\
& {\textsc{LXP10+20}} & $6.65^{+1.09}_{-0.92}$ & $0.22^{+0.02}_{-0.02}$  & $0.13^{+0.026}_{-0.024}$ & $0.12^{+0.01}_{-0.01}$ & $0.98^{+0.28}_{-0.20}$ & $0.097^{+0.010}_{-0.011}$ & $0.107^{+0.002}_{-0.002}$ & 76.43/63.00=1.21\\
& {\textsc{SXT}} & $7.34^{+2.03}_{-1.51}$ & $0.16^{+0.03}_{-0.03}$  &   $0.13(f)$ & {$<$~0.05} &   ---- &   ---- & $1.63^{+0.09}_{-0.09}$ & 114.32/140.00=0.82\\

\textbf{2019} & {\textsc{LXP20}} & $11.40^{+1.07}_{-1.08}$ & $0.13^{+0.01}_{-0.01}$  & $0.17^{+0.02}_{-0.02}$ & $0.821^{+0.006}_{-0.06}$ & $1.19^{+0.12}_{-0.10}$ & $0.11^{+0.005}_{-0.005}$ & $0.0280^{+0.0005}_{-0.0005}$ & 114.80/99.00=1.16\\

& {\textsc{SXT}} & $10.95^{+2.56}_{-1.87}$ & $0.13^{+0.02}_{-0.02}$ & $0.17(f)$ & $0.107^{+0.063}_{-0.064}$ & ---- & ---- & $0.598^{+0.034}_{-0.034}$ & 44.760/45.000=0.997\\
\hline \hline
\end{tabular}
\begin{flushleft}
Notes : $\nu_{L1}$ is the  break frequency, $\nu_{Brk}$. The fixed parameters are shown with '(f)'. \\
\end{flushleft}
\label{tab:PDS_lorentz}
\end{table*}

In order to study the variability of the source GX 339-4, we utilized the broadband Power density spectrum. For the timing analysis of the source we created the PDS using the HEASoft tool POWSPEC \url{https://starchild.gsfc.nasa.gov/xanadu/xronos/help/powspec.html}. For 2017 observation, the PDS was generated for  LAXPC10 and LAXPC20 individually and combined. In POWSPEC, the PDS was created from 0.1~s binned background subtracted lightcurve in the energy range 3.0-20.0~keV. The lightcurve was divided into 35 segments with length of each segment as 1.64~ks. Geometric rebinning was undertaken to smoothen the PDS. These segments were then averaged producing the final PDS of the observation. Also, the normalization is such that the PDS can be expressed in units of ${rms}^2/Hz$. Similarly for 2019 observation, the PDS was generated from 0.1~s binned background subtracted lightcurve in the energy range 3.0-20.0~keV for LAXPC20 detector only. The lightcurve was divided into 44 segments with length of each segment to be 1.64~ks. Geometric rebinning was undertaken to smooth out the PDS. Similar to 2017 observation, the normalization is such that the PDS can be expressed in units of ${rms}^2/Hz$. 

Further, the PDS of the source with SXT data was generated from 2.3775~s binned (since it's the minimum time-bin value for SXT) background subtracted lightcurve in the energy range 0.6-7.0~keV. The lightcurve was divided into 44 and 59 segments for 2017 and 2019 observations respectively. Each segment was of length 0.512~ks. Geometrical rebinning was considered to smoothen the PDS. These segments were then averaged producing the final PDS of  each observation.

The PDS of both observations showed no sharp features but a broadband noise continuum with a low-frequency break, which is typical for a low luminosity hard state. Therefore for fitting the broad components we employed a combination of few zero-centered Lorentzians, given individually by,
  
\begin{equation}
  {L(\nu)}~=~\frac{N_L \Delta}{\pi} 
              \left(\frac{1}{(\nu-\nu_o)^2+\Delta^2}\right) 
\label{eq:lor}
\end{equation}

\noindent with $\nu_o$, $N_L$ and $\Delta$ as its centroid frequency (which is 0 for zero-centered Lorentzian), normalization and HWHM respectively \citep{belloni2002unified}. We expressed each Lorentzian component with its characteristic frequency,  $\nu_{L}$~\(=\sqrt{\nu_o^2+\Delta^2} = \Delta\), also known as the peak frequency i.e. the frequency at which $\nu~P_\nu$ peaks. We also implemented a constant to fit the Poissonian noise ($N_{noise}$).

We applied the above model to the PDS and found that that three zero-centered Lorentzians were sufficient to describe the LAXPC PDS. However, only two Lorentzian components were required to fit the SXT data, since the third component used for the LAXPC PDS fitting, peaked at a frequency higher than the SXT frequency range. The peak frequency of the first Lorentzian component ($\nu_{L1}$) in each fit is identified as the low-frequency break ($\nu_{Brk}$). Additionally, the peak frequency of the second component $\nu_{L2}$ for SXT was fixed at the value obtained from the LAXPC fit. Since the fit parameters for LAXPC10 and LAXPC20 data were found to be similar, we
jointly fitted the two for the 2017 observation.  The fitting of $\nu~P_\nu$ curves for both observations is shown in Figures \ref{fig:PDS_lorentz_2017} and \ref{fig:PDS_lorentz_2019}. We plot the noise subtracted PDS for better understanding of the peaking components. For SXT, 2017 the PDS is shown upto 0.025 Hz as above these points the Poisson noise dominated the PDS. The best fit values of parameters are mentioned in Table \ref{tab:PDS_lorentz} along with the reduced $\chi^2$.

In Figure \ref{fig:Brfq_flux}, we plot this peak frequency of the first Lorentzian component (which we also refer to as the "break" frequency) as a function of the unabsorbed flux in the 3.0-9.0~keV band. We compared our results with other detections done in similar low/hard state and  reported their unabsorbed flux values mentioned for energy range 3.0-9.0~keV and in case it was not mentioned we made use of reported spectral parameters to obtain the flux values in desired energy range. We compared our results with those obtained by \cite{plant2015truncated}, who have reported variation of break frequency with the unabsorbed flux for the low/hard state observation of GX 339-4. The plot  shows that the results obtained in this work are consistent with the trend observed by \cite{plant2015truncated}. We also plot other detections quoted in different works for \textit{e.g.} \cite{nandi2012accretion} have also reported break frequency for six observations of which one of them is in the hard state, while the others are in the soft or intermediatory states. For their only hard state observation, we found that the break frequency is significantly higher than the value reported by \cite{plant2015truncated} for similar flux levels, but it may be noted that their observation was just before a transition to the soft state and had a relatively high inner disk temperature of $\sim 1.3$~keV.  The break frequencies and flux values reported by \cite{stiele2015energy} are closer to the values obtained by \cite{plant2015truncated}. \cite{migliari2005} have also presented estimates of the break frequencies for GX 339-4, however they do not tabulate the spectral parameters. Therefore we report the absorbed flux levels mentioned in \cite{corbel2003} for 3.0-9.0~keV band for the same observations, it should be mentioned that at $>$3~keV these flux levels should not be much different from the unabsorbed flux levels. Their values are found to be consistent with those obtained by \cite{plant2015truncated} and this work. Also, \cite{stiele2017nustar} report the break frequencies in the hard state of GX 339-4 but use a complex model to fit the energy spectrum therefore we estimated unabsorbed flux levels in 3.0-9.0~keV with spectral parameters reported for the same observations analysed in \cite{wang2018}. Figure \ref{fig:Brfq_flux} shows these values quoted from literature along with the AstroSat results.

\section{Summary and Discussion}
\label{sec:Summary_Discussion}

We have analysed the black hole candidate GX 339-4 when it was at the beginning of its outbursts in 2017 and 2019. For 2017 observation, the joint spectral analysis of LAXPC and SXT showed that an absorbed power-law could well describe the photon spectrum with photon index, $\Gamma$~=~1.57 and column density, $N_H$ ($10^{22}$ $cm^{-2}$)~=~0.43. It was in faint low/hard state as the unabsorbed flux in energy range 3.0-9.0~keV was only 0.85~$\times$~$10^{-10}~erg~s^{-1}~cm^{-2}$. The spectrum consisted of weak reflection features such as Fe k$\alpha$ line and beginning of Compton hump, therefore modelling the reflection emission with two different reflection models  \textit{Ireflect} and \textit{Relline} improved the $\chi^2$ fit. XSPEC model \textit{Ireflect}, assumes reflection from an ionized accretion disk whereas \textit{Relline} applies the relativistic effects to the iron line. Similar results were found for 2019 observation as the photon spectrum was modelled by an absorbed power-law with $\Gamma$ $\sim1.58$ and $N_H$ ($10^{22}$ $cm^{-2}$)~=~0.44. This too was a faint low/hard state observation but with flux $\sim4$ times higher (3.44$\times$ $10^{-10} erg~s^{-1}~cm^{-2}$) than the 2017 observation in the same energy band. This observation was also modelled with reflection components \textit{Ireflect} and \textit{Relline} leading to improvement in $\chi^2$ fit.

\begin{figure}
\centering
\caption{Variation in break frequency (mHz) with unabsorbed flux ($10^{-10} erg~s^{-1}~cm^{-2}$) in energy range 3.0-9.0~keV as plotted for this work, Plant et al. 2015 and for values obtained by Migliari et al. 2005, Nandi et al. 2012, Stiele et al. 2015 and Stiele et al. 2017.}
\vspace{0.5cm}

\includegraphics[height=7cm, width=1.0\linewidth]{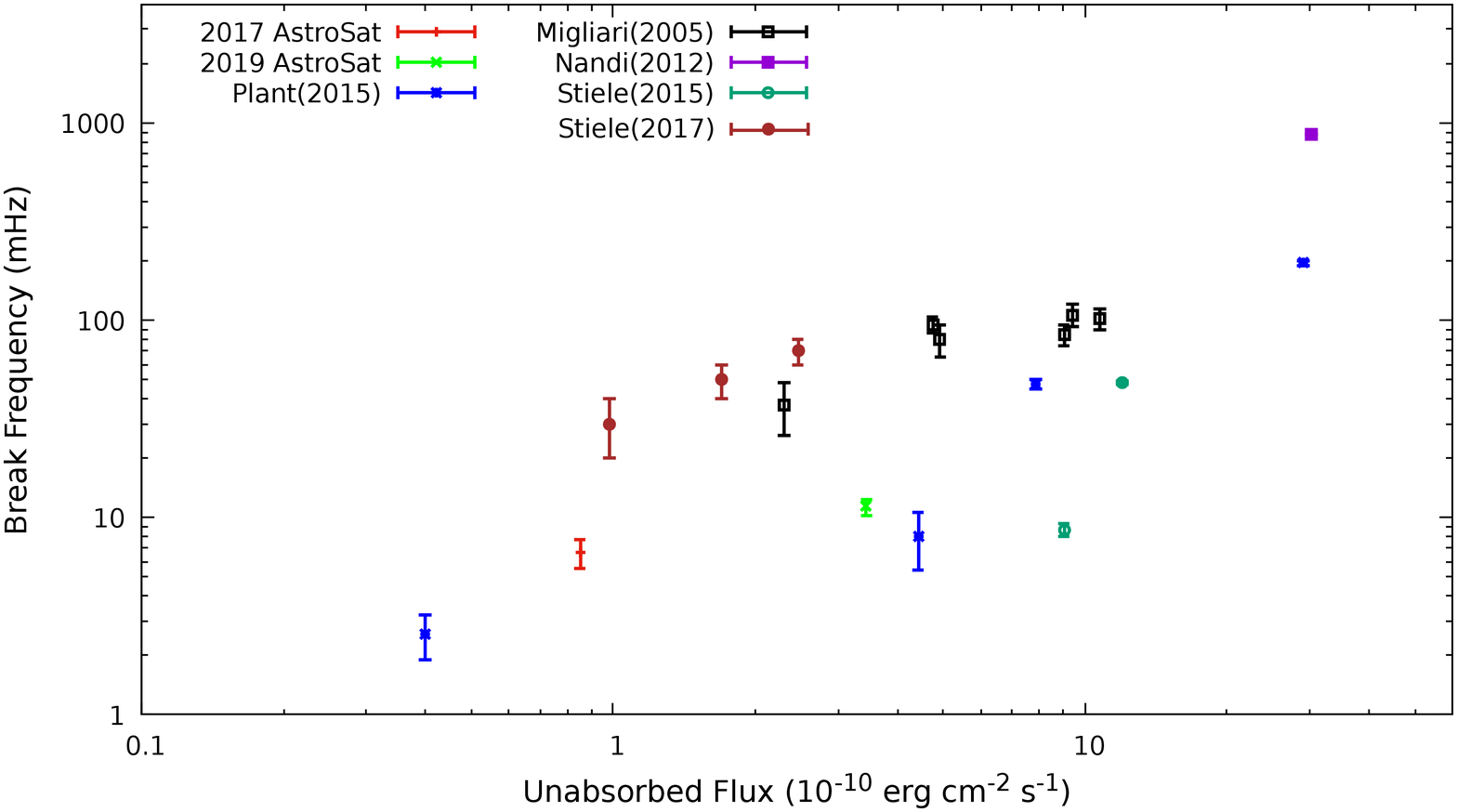}
\label{fig:Brfq_flux}
\end{figure}

The timing study of GX 339-4 was accomplished using the Fourier transformation technique. The PDS were fitted with a multi-Lorentzian model and the fitting identified a mHz break for each of the observation.  The 2017 observation showed a break at  $\sim6~mHz$  for LAXPC20.  The detection of this mHz break is validated from three independent detectors LAXPC10, LAXPC20 and SXT. The 2019 observation showed a break at $\sim11~mHz$ for LAXPC20 and its detection is also confirmed from two different detectors LAXPC20 and SXT, consistent to each other within errorbars. It can be clearly seen that the break for 2019 observation is detected at a higher frequency than 2017 observation. Detection of breaks at such low flux is in accordance with the result obtained by \cite{plant2015truncated} where RXTE observations of GX 339-4 in hard (rise) as well as hard (decay) states exhibited mHz break frequencies in the PDS. The unabsorbed flux level (3.0-9.0~keV) for these observations was in the range 0.4~-~29.0 $\times$ $10^{-10}~erg~s^{-1}~cm^{-2}$. They used the XSPEC model \textit{Phabs*Powerlaw} to fit the photon spectrum and found the photon index to be 1.565, which is also closer to the photon index we have found for our observations. Also, these observations show a trend in break frequency moving towards higher frequency from {2.55} to {195} mHz with the increasing flux. As mentioned in section \ref{sec:Timing}, we also include points from previous hard  state observations of GX 339-4 (Figure \ref{fig:Brfq_flux}) and find that our values roughly fit in the evolution of break frequency with flux as seen in different observations of GX 339-4.
 
This evolution of break frequency is according to the truncated disk model (\cite{done2007modelling, ingramviscous2011} and references therein). In this model, low L/$L_{edd}$ state has a truncated accretion disk at a certain radius ($R_{in}$) after which the disk is replaced by a geometrically thicker and optically thinner hot inner flow (or corona) \citep{rozanska2000two,mayer2007time}. The spectral transitions in this model are attributed to the modification of the geometry of accretion disk and corona. Also, this model associates the characteristic frequencies in the PDS with frequencies of the two flows.  For \textit{e.g.}, the lowest characteristic frequency ($\nu_{brk}$) in the PDS of black hole binaries is associated with the viscous time scale \citep{churazov2001soft,gilfanov2005} at the truncation radius of the accretion disk (or the outer edge of the hot inner flow), which is since at the truncation radius of the disk the viscous time scale is much greater than the thermal or dynamic time scales \citep{frank2002accretion}. With the source evolving to higher luminosity states, the truncation radius approaches the ISCO. Therefore, with this geometrical change all characteristic frequencies in the PDS are expected to move to higher frequencies. Figure \ref{fig:Brfq_flux} shows a similar behaviour where the break frequency is seen to evolve with the flux. In addition to this \cite{plant2015truncated} have observed the evolution of truncation radius with the luminosity which has been also observed in \cite{tomsick2009truncation} and many other works which is also in line with the truncated disk model.

As already mentioned above,  the  characteristic low-frequency break can be associated with the viscous time scale ($t_{visc}$) at the truncation radius of the accretion disk and the expression for $t_{visc}$ for a standard accretion disk  \citep{frank2002accretion, done2007modelling} is given by,

\begin{equation}
  {t_{\rm visc}}~=~4.5\times 10^{-3} \frac{1}{\alpha} 
              \left(\frac{H}{R}\right)^{-2} 
               \left(\frac{R}{6~R_g}\right)^{3/2} \left(\frac{M}{10~M\textsubscript{\(\odot\)}}\right)
\label{eq:tvisc}
\end{equation}

\vspace{0.3cm}
\noindent Here, $\alpha$ is the dimensionless viscosity parameter as prescribed in \cite{shakura1973}, $H/R$ is the semi-thickness of the disk and M is the mass of the compact object. For obtaining the estimate on truncation radius of the inner disk, we used the values of break frequency obtained from the LAXPC20 PDS fitting ($\nu_{L1}$) for both observations. We associated these frequencies to the viscous timescale and estimated the inner radius as prescribed in the expression \ref{eq:tvisc} with $\alpha$~=~0.1 and H/R~=~0.1 (parameters for a thin accretion disk) and mass of black hole~=~6\(M_\odot\). The truncation radius was found to be at 92.5$\pm$3.8~$R_g$ and 61.08$\pm$1.42~$R_g$ for 2017 and 2019 observation respectively. The error bars represent only the uncertainty in the low-frequency break measurement and do not take into account that the values of $\alpha$, $H/R$ and black hole mass may be quite different than the ones assumed. Thus, while the $R_{in}$ estimations are approximate, the results do show a higher truncation radius for the lower flux 2017 observation. Moreover, these values are consistent with our assumption during spectral fitting that its value is $\sim$100~$R_g$. Estimation of the truncation radii can also be made by detailed spectral fitting, but they depend on the specific model used. For example, for the same lowest flux observation, \cite{plant2015truncated} report values of 79, >321 and >344 $R_g$ when they used different spectral models. Nevertheless, spectral analysis support the general trend that the truncation radii increases with decreasing luminosity \citep{petrucci2014,wang2018,garcia2015,garcia2019}.\\

To conclude the results, we observed mHz breaks in the PDS of GX 339-4 for two different observations of its faint low/hard state. We confirm these mHz detections with independent detectors (LAXPCs and SXT). These breaks are also few of the lowest frequency breaks observed for this source which may have a connection with very low flux of these observations. The comparison of these low-frequency breaks with breaks observed for few other low/hard state observations imply that both our observations nearly fit in the evolution of the break frequency with source flux. This evolution along with the trend of inner disk radius moving closer to the ISCO with the increased luminosity agrees well with the truncated disk model. However, the spectral fitting shows that few spectral parameters were not constrained properly, therefore modelling data with better spectral resolution data such as from NuSTAR can result in finding better constraint on these parameters including the truncation radius. Also, long exposures of the source with for \textit{e.g.} SWIFT/XRT or NICER could provide better spectral modelling.

\section{Acknowledgement}

We are thankful to the anonymous referee for the valuable comments which improved this work.  NH is grateful to the LAXPC and SXT teams for providing the data and requisite software tools for the analysis. NH is also grateful to the Inter-University Centre for Astronomy and Astrophysics (IUCAA) for allowing frequent visits to work on this project. NH would like to thank Akash Garg and Sneha Prakash Mudambi for their discussions related to LAXPC and SXT data analysis. NH acknowledges the funding received from the Department of Science and Technology (DST) under the scheme of INSPIRE fellowship.

\section{Data Availability}
AstroSat data analysed in this work can be accessed through the Indian Space Science Data Center (ISSDC) website (\url{https://astrobrowse.issdc.gov.in/astroarchive/archive/Home.jsp}). The data was analysed with the packages of HEASoft version (6.26.1)  (\url{https://heasarc.gsfc.nasa.gov/docs/software/heasoft/}). The softwares used for LAXPC and SXT data reduction are available on \url{http://astrosat-ssc.iucaa.in/?q=laxpcData} and \url{http://www.tifr.res.in/~astrosat$\_$sxt/dataanalysis.html}, respectively. 


\bibliography{References}

\end{document}